\documentclass[a4paper,11pt]{llncs}

\usepackage[english]{babel}
\usepackage{url}

\begin{document}

\author{Fabian van den Broek}
\institute{Radboud University, Nijmegen \\ Institute for Computing and Information Sciences (iCIS)}
\title{Eavesdropping on GSM: state-of-affairs}
\maketitle

\begin{abstract}
In the almost 20 years since GSM was deployed several security problems have been found, both in the protocols and in the - originally secret - cryptography. However, practical exploits of these weaknesses are complicated because of all the signal processing involved and have not been seen much outside of their use by law enforcement agencies.

This could change due to recently developed open-source equipment and software that can capture
and digitize signals from the GSM frequencies. This might make practical attacks against GSM much
simpler to perform.\\

Indeed, several claims have recently appeared in the media on successfully eavesdropping on GSM. When looking at these claims in depth the conclusion is often that more is claimed than what they are actually capable of. However, it is undeniable that these claims herald the possibilities to eavesdrop on GSM using publicly available equipment.

This paper evaluates the claims and practical possibilities when it comes to eavesdropping on GSM, using relatively cheap hardware and open source initiatives which have generated many headlines over the past year.
The basis of the paper is extensive experiments with the USRP (Universal Software Radio Peripheral) and software projects for this hardware.
\end{abstract}


\section{Introduction}
GSM was developed in the late 1980s and deployed in most Western countries in the early 1990s. Since then GSM has seen an enormous rise both in its coverage and in the number of subscribers.

GSM is perhaps the most successful technology of the last twenty years. A survey by the International Tellecommunication Union (ITU) showed that by the end of 2008 around 1.5 billion people in the world (some 23\%) use the Internet. But there were around 4.1 billion people in the world (over 60\%) who had a mobile subscription, while over 90\% of the worlds population lived in a region that at least has access to GSM \cite{guardian}.
These are staggering numbers showing a tremendous spread of GSM technology.\\

The fact that GSM has weaknesses is nothing new. The lack of mutual authentication in GSM -- a mobile phone authenticates itself to the cell tower, but not vice versa -- was quickly seen as a problem \cite{anderson:secEng}. Also GSM specifically does not use point to point encryption between callers. It only encrypts the messages while on the air interface. This allows law enforcers to tap conversations in the core of the GSM network.

Meanwhile more and more services are being deployed on top of the GSM network, increasing the incentive for criminals to attack GSM. In several countries you can pay for services or products via text messaging. Several Internet banking applications use the mobile phone as an external (out-of-band) channel to verify transactions. The Dutch ING bank stated only last January that they will start to use the mobile phone to send users their password reminders, even though they already use it for transaction verification \cite{url:ING}.
Where previously making un-billed calls was the major economic attraction in attacking GSM, increasingly real money can be made.\\

At the end of 2009 some pretty large tables usable for a brute-force attack against the main cipher used in GSM where released. This has let to many wild claims on the insecurity of GSM in the media.
This document examines the feasibility of these claims, based on experimental work with a USRP -- an open-hardware generic radio transceiver -- and several open-source software products, which could, in theory, be used to eavesdrop on GSM.

At about the same time some practical examples of Man-in-the-Middle (MITM) attacks on GSM surfaced. These are possible because in GSM the cell towers do not authenticate themselves to the mobile phones. So it is possible to act as a cell tower towards a mobile phone. Currently these attacks lack the ability to act as a cell phone towards a cell tower, so only a limited form of a MITM attack has been shown, in which an attacker acts as a genuine cell tower, instructs the cell phone to not use encryption, and transfers outgoing calls via a VOIP connection. This attack can only capture calls being made from the cell phone under attack, and not incoming calls. This document will limit itself to eavesdropping, and thus excludes these MITM attacks. 

Section \ref{sec:equip} discusses the equipment that was used during our experiments. Section \ref{sec:theo} will discuss the theoretical steps required to passively eavesdrop on GSM, while Section~\ref{sec:prac} will discuss the current status of several practical projects implementing these steps. Section~\ref{sec:cm} discusses some possible countermeasures against an eavesdropping attack.

\section{Equipment used}\label{sec:equip}
In order to evaluate the practicality of eavesdropping attacks, we experimented with some hard and software. Specifically this were:
\begin{itemize}
\item The USRP,
\item combined with the DBSRX daughter board
\item and running GNU Radio and AirProbe.
\end{itemize}
\begin{itemize}
\item A Nokia 3210 mobile phone
\item connected to a pc running the Gammu software.
\end{itemize}
The USRP (Universal Software Radio Peripheral) is a general purpose, open-hardware, transceiver that can be linked to a computer via USB, and handles the receive part of an eavesdropping attack. The USRP, and its successor, USRP2, are discussed in more detail in Section \ref{sec:prac_capt}.

The USRP has to be extended with a daughter board in order to receive the correct frequency spectrum. We used the DBSRX daughter board in this research, which is a 800MHz to 2.4GHz receive-only board, covering all basic GSM frequency bands.

The USRP was controlled using the GNU Radio software and the AirProbe software on top of that. Both software products are detailed in Sections \ref{sec:prac_capt} and \ref{sec:prac_int}.\\

Next to this we also made extensive use of a Nokia 3210 GSM phone, connected to a computer, also via USB, running the open-source Gammu \cite{url:Gammu} project. This combination enabled us to force the Nokia 3210 in a debug mode that transparently logs all packets sent to and from the phone. 

The Gammu + Nokia phone method has much better reception than the USRP + AirProbe, after all the mobile phone is specifically made to receive these signals. Though you can only see the messages to or from the specific phone connected to the computer. You cannot see any message for other phones, nor is it possible to change the phone's behavior in this. So this is a great practical aid to get a better grasp of the GSM protocol and to finetune the USRP, but it lacks the versatility to be useful in a eavesdropping attack.

\section{How to eavesdrop on GSM in theory?}\label{sec:theo}
Eavesdropping on GSM, or probably any communication system for that matter, can be broken down into three stages:
\begin{enumerate}
\item \label{cap} Capturing the signals.
\item \label{dec} Decrypting the captured signals.
\item \label{int} Interpreting the decrypted signals.
\end{enumerate}
This section will look at these steps in more detail. Section \ref{sec:prac} will look at the current state of practical open-source projects usable for these steps.

\subsection{Capturing the signals}\label{sec:th_capt}
Already the first stage has been a major obstacle for many years. Specialized equipment to capture the GSM signals has long been very expensive and often proprietary. GSM can be used on several frequency bands, but most commonly used are the GSM-900 and GSM-1800 bands. The frequency bands are divided into channels of 200 KHz wide each. In a typical conversation two of these channels will be used at any given time for a mobile phone to communicate with the cell tower; one channel for each direction. These channels are separated by a constant offset.

One part of user data in a GSM network, e.g.\ 20ms of speech data, is transmitted in four packets, called \textit{bursts} in GSM. These bursts are modulated radio waves transmitted in a time slot of 576.9 $\mu$s. 
Most GSM networks employ \textit{channel hopping}, which is used as a signal quality measure, and causes the transmission to switch to a new channel after every single burst.
The challenge in capturing the GSM signals lies in receiving the bursts on time and in demodulating them correctly.

Section \ref{sec:prac_capt} discusses projects and hardware that can be used to capture GSM signals.

\subsection{Decrypting the captured signals}\label{sec:th_dec}
Assuming that encryption is enabled on the GSM network, decrypting the captured bursts is the next step. There are three encryption algorithms defined for GSM; A5/1 and A5/2, both stream ciphers, and A5/3, a block cipher. Of these three A5/2 is by far the weakest and can be broken in less than a second on a personal computer with only a few dozen milliseconds of cipher text \cite{barkan:a512}. The A5/3 algorithm is considered the strongest encryption of the three. A5/3 very recently saw a theoretical break \cite{dunkA53}. This attacks requires 2$^{26}$ chosen plain text messages encrypted under related keys. For now this does not lead to a practical attack on A5/3, though it is cause for concern since this weakness does not exist in MISTY, the cipher that A5/3 was based on. At the moment of writing no attack on A5/3 is feasible, the future will tell whether this remains so.

A5/1 is the main encryption algorithm used in most Western countries. It is a stream cipher with three registers that clock irregularly and have a combined size of 64 bits --- which, conveniently, is also the size of the session key. The session key is computed by the SIM card and in the provider's home network from a random challenge and a secret key by an undisclosed proprietary algorithm. Originally most providers used an algorithm called COMP128, which was reverse engineered in 1998 by Briceno et.\ al.\ \cite{Briceno:a51}. This showed that COMP128 actually delivered a 54 bits session key with ten appended zeros. It is unknown which algorithms are currently used by providers to generate the session key and if the session keys are still weakened.

The A5/1 algorithm was actually the first encryption algorithm used in GSM. It was originally kept secret and was only disclosed to GSM manufacturers under an NDA. In 1999 though, Marc Briceno reverse engineered the design of both A5/1 and A5/2 from a GSM phone \cite{Briceno:a51}. Several attacks against A5/1 have been published since then \cite{golicA51,barkan:a512,barkanA51}.

Recently a practical implementation for a brute-force Time-Memory-Trade-Off attack against A5/1 was announced, which will be discussed in Section \ref{sec:prac_dec}. This attack is in fact an improved version of an attack proposed by Elad Barkan et.\ al.\ in 2003 \cite{barkan:a512}.

\subsection{Interpreting the decrypted signals}\label{sec:th_int}
After the signals have been captured and deciphered they still need to be interpreted. The payload of the bursts needs to be reordered and can be checked for transmission errors. Besides the cryptography, all the specifications of GSM are public, so this does not require any reverse engineering.

Several projects that implement a GSM stack are discussed in Section \ref{sec:prac_int}

\section{The open-source practical implementation}\label{sec:prac}
Eavesdropping on GSM is possible --- equipment can be bought that allows for eavesdropping on A5/1 and A5/2 encrypted conversations \cite{url:eavesdropper}, but this equipment is sold restrictively to law enforcement agencies and military and at a high price. So it is not a question on whether it is possible to eavesdrop on GSM, but rather how hard it is using publicly available hardware.

Recently several open-source projects have surfaced that either aim to prove the insecurity of GSM, or simply offering open-source implementations of core GSM network elements.
So how far are the practical realizations of the three steps from the previous section in the open-source community? For each of these steps there is a possible open-source project to use. Respectively these are: 
\begin{enumerate}
\item \label{capi} USRP together with GnuRadio and AirProbe, usable for signal capturing
\item \label{deci} Kraken, for the decryption of the captured signals
\item \label{inti} OpenBTS or OpenBSC or AirProbe, to interpret the messages
\end{enumerate}
Because the AirProbe project is actually a collection of tools usable both for capturing and for the interpreting of the GSM bursts, the project is listed twice. We will now look at each of these projects in more detail.

\subsection{Capturing using USRP together with GnuRadio and AirProbe}\label{sec:prac_capt}
Capturing GSM signals can be done using the USRP and GNU Radio combination.
The Universal Software Radio Peripheral (\textit{USRP}) is an open-hardware device developed by Matt Ettus and which can be ordered through his company Ettus Research \cite{url:ettus}. It is a transceiver that can be linked to a computer and can be tailored to specific frequencies by extending it with daughter boards and attaching the appropriate antenna. The USRP contains a programmable FPGA which can be used to perform some signal processing. In its standard configuration a USRP creates 16 bit I and Q samples when receiving a given frequency. These are complex samples, with the real part (Q) describing the cosine of the signal, and the imaginary part (I) describing the sine of the signal plus 90 degrees. One sample is thus 32 bit long and can be sent to the host computer through the communication port, for further processing.

There are currently two types: the USRP and the USRP2. The USRP (or USRP1) can receive a bandwidth of 32MHz and can transmit on a bandwidth of 64MHz. It transmits the samples to the host computer via a USB2.0 connection, which has a practical maximum data throughput of 32 Mbyte/s. The USRP2 can receive a bandwidth of 50 MHz and transmit on a bandwidth of 200 MHz wide. Compared to the USRP1, the USRP2 also contains a much faster FPGA and a Gigabit Ethernet port instead of the USB connection. 

The USRPs have a 64MHz crystal oscillator internal clock, while most GSM phones use a 13MHz symbol clock with a much better accuracy. Of course the 64MHz samples can be re-sampled to (a multiple of) 13MHz, although this brings an extra computing complexity.  Also the USRP's oscillators are much less accurate and can show quite some drift when compared to GSM system clocks, resulting in bad reception. An external clock can be attached to the USRPs. Using a more accurate external clock providing a pulse at (a multiple of) 13MHz solves these issues.\\

\textit{GNU Radio} \cite{url:gnuradio} is a free software toolkit licensed under GPL for implementing software-defined radios. It was started by Eric Blossom. It works with several different types of RF hardware, such as soundcards, but it is mostly used in combination with an USRP. Basically GNU Radio is a library containing lots of standard signal processing functions, such as filters and (de)modulations.\\

GNU Radio, out-of-the-box, does not offer much in terms of GSM sniffing capabilities. However, GNU Radio can be used by other software packages, such as \textit{AirProbe}, to perform the low level functions of GSM sniffing, such as reception and demodulation. AirProbe \cite{url:airprobe} is an open-source project trying to built an air-interface analysis tool for the GSM (and possible later 3G) mobile phone standard. One part of the project handles the reception of GSM signals (using the GNU Radio functions) while another part can also be used to interpret the GSM signals, which is why we will get back to AirProbe in section \ref{sec:prac_int}.
Currently AirProbe is only able to listen to the down link (cell tower $\rightarrow$ mobile phone) of conversations, so some development is still required.\\  

The main problem for reception is the channel hopping used in most GSM networks. 
Using the AirProbe software out of the box helps you receive a single carrier on the down link (messages from cell tower to mobile). In order to capture an entire conversation when channel hopping is used, you will need a way to gather all the bursts on all the different frequencies. There are two general approaches to achieve this:
\renewcommand{\labelenumi}{\Roman{enumi}.}
\begin{enumerate}
\item \label{it:follow}Let the USRP follow the hopping sequence.
\item \label{it:all}Capture all possible frequencies and attempt to follow the sequence afterwards. 
\end{enumerate}
\renewcommand{\labelenumi}{\arabic{enumi}.}
Approach I requires a lot of processing inside the USRP's FPGA. All the parameters for the hopping sequence need to be retrieved from certain bursts, then the hopping sequence needs to be calculated and followed for every burst. There are three possible configurations for mobile networks in which to transmit the hopping sequence parameters. In one of these the parameters are transmitted in the clear. In the other two configurations these parameters are transmitted after encryption has been enabled. During our experiments we only ever observed cell towers that use so-called ``early assignment'', in which the hopping parameters are transmitted under encryption. Besides, the network can always command a new hopping sequence under encryption, irrespective of which configuration is used. This necessitates breaking the encryption really fast in order to follow the hopping sequence in time. This is currently not possible. The USRP's FPGA only samples a certain frequency at a certain rate and sends these samples to a computer. So having the FPGA decrypt and interpret the bursts in order to follow the hopping sequence will require a lot of implementation. It is questionable whether even the USRP2's much faster FPGA will be able to decrypt messages and then compute the hopping sequence in time. This approach might need additional FPGAs, and thus additional costs, to pull off. Also the tune delays of the USRP's hardware (the time between a ``tune'' command and the moment the USRP retrieves usable samples from the desired frequency) seem to large at the moment and need to be brought down. However, it is an approach that should work for every cell tower and regardless of the amount of traffic.\\

Approach II requires the capture of large amounts of data, namely to log all channels and determine the hopping sequence later. The problem here lies in reducing the data the USRP sends on to the computer. A hopping sequence can maximally hop between 64 carrier frequencies. These carriers are all 200 KHz wide, and can be spread out evenly on the entire GSM spectrum. So worst case the attacker has to capture the entire GSM band (for GSM900 this is 25MHz for the up or down link). The problem here is data throughput to the PC. Each sample of an USRP is represented by two 16 bit numbers. For the USRP1's USB2.0 connection this means that the maximum bandwidth that can be sent to the computer is around 8MHz (8MHz $\times 2 \times 16 = 256$Mbit/s). The USRP2's GBE connection can manage around 30MHz, which is enough for one sided capture of GSM900. This would require the host computer to be able to process 100 MByte/s of data (25MHz $\times 2 \times 16$ = 800Mbit/s) and even 200 MByte/s for both up and down link, which is too much for most PCs.
 
Of course some optimizations are possible. A single cell tower never serves the entire GSM frequency band. This means that you can already discard all carriers above the top frequency and all carriers below the lowest frequency of a specific cell tower. However, that approach will not work for most situations, since the maximal number of carriers (64) for a single cell tower is still too much for the USRP. Also, since the USRP can only receive a continuous frequency band, if the top and bottom frequencies are too far apart, this approach will not work.

Another optimization would be to have the FPGA discard all channels that have no traffic on them. This will of course only be effective if only a few phones are active --as in calling -- at the same time. It would be even better to have the FPGA interpret enough of the bursts, so it can already drop some that are not a part on the conversation the attacker tries to capture. This optimization does see the same problems as those with the first approach because it requires a lot of FPGA computation -- though less so, because the FPGA does not need to crack A5/1 in the second approach.

The objections stated above, however, do not form a problem if the cell tower under attack does not employ channel hopping, or only transmits on a few frequencies in a tight spectrum. On such cell towers eavesdropping using an USRP seems a genuine possibility. It is not clear how many, if any, of the cell towers in operation, match one of these conditions. This makes it hard to estimate the risk in the current situation. During this research only a handful of cell towers were observed, but none of those fulfilled these conditions that would allow eavesdropping using the USRP.\\

Currently approach II seems to be the one that most people in the AirProbe community believe is the correct one to follow. It does indeed seem the easier way out, with a higher chance of success (on at least some cell towers), though the FPGA programming needed here is by no means a simple task. At least no one has communicated a way to tackle this. This approach will probably not work in every situation, given the problems with data throughput to the PC, but a working implementation for some cell towers is enough for the AirProbe community to show that they can listen in on GSM.\\

\subsection{Decrypting using Kraken}\label{sec:prac_dec}
A project was publicly announced in August of 2009 suggesting a way to efficiently break the A5/1 cipher. This project runs under the, slightly unimaginative, name \textit{A5/1} \cite{url:a51}. However, in July 2010 the look-up tool for this project was released and named \textit{Kraken}. To avoid any confusion with the cipher we will refer to the A5/1 project by the name Kraken.

The Kraken project mainly consists of creating large tables in a generic time-memory trade off. This had been proposed before \cite{hulton:A51}, but the distinguishing factor of this new project is that instead of computing the tables at a single point everybody on the Internet can join in and compute a table and then share them via bit torrent \cite{url:torrents}. The code to compute these tables can be downloaded and it runs on certain types of NVIDIA and ATI graphics cards.\\

The idea behind these tables is as follows. The contents of several bursts that are sent through the air, after encryption is enabled, can for the most part be guessed. This gives known plain text samples. XORing those plain text samples with the actually received cipher text reveals keystream samples. The tables now function as a code book with 64 bits of keystream that are mapped to internal A5/1 states producing that exact piece of keystream.

There are $2^{64}$ possible internal states for A5/1. That number is too large to be able to map all internal states. So instead of just storing an internal state and its 64 bits of keystream, they actually compute large chains where for each link 64 of the output bits are again used as the internal state. They only store the begin and end points of these chains. Now when a piece of keystream is recovered, the attacker starts to make a chain out of this keystream, but for every link he checks whether the resulting value is stored in its table. If the attacker finds a hit, then the attacker can recover the internal state by computing the original chain from its stored begin point -- from the internal state the session key that was used can be computed, allowing for decryption of the entire communication. This shrinks the size of the tables, but the attack time is increased and the tables no longer guarantee that an internal state can be found.  
But a much bigger problem that surfaces with this approach is chain mergers. Several different internal states will compute the same 64 keystream bits causing different chains to `merge' and cover the same part of the key space. This makes the tables much less effective. To counter some of these problems a combination of two techniques, one to decrease the attack time and one to decrease the number of chain mergers, is used. Those techniques are distinguished points and rainbow tables respectively. Combining these techniques has been proposed before \cite{Erguler:tmto,Hong:dp_rbt}.


Because of the chain mergers, a certain number of different tables need to be produced. Those tables also need to cover a certain amount of the entire key space. Recently a set of tables, named the `Berlin set' was released together with the lookup tool, Kraken. The tables are distributed via bittorrent in a special transport format. Currently this is around 1.5TB of data. This transport format can then be transcoded into the actual read format, which takes just under 1.7TB disk space. Currently the tables cover around 22\% of the keyspace. If the key is in the tables, then Kraken will typically find it within a few minutes (around 1 to 4 minutes) on a Intel Core2 Quad 2.33GHz machine with the tables divided over several disks. Using solid state memory instead of conventional hard drives for the table storage should significantly improve on this look up time, at the expense of an increase in the financial costs.

A major drawback of this approach is that it requires faultless reception of 64 consecutive bits. A single unnoticed error in a reception will make it impossible to retrieve the session key using these tables.
In GSM the encryption is performed on top of the error detection codes, so the error detection can not be used to find reception errors before the decryption step. Currently the reception from the USRP, while running AirProbe, does not provide the faultless reception needed by Kraken. 

\subsection{Interpreting using OpenBTS, or OpenBSC, or AirProbe}\label{sec:prac_int}
After the demodulation and decryption steps the bursts need to be interpreted. 
There are currently several open-source projects that implement at least part of the GSM stack. These are the OpenBTS\cite{url:openbts}, OpenBSC\cite{url:openbsc} and AirProbe\cite{url:airprobe} projects.

The \textit{OpenBTS} and \textit{OpenBSC} projects both aim to offer a functional open-source GSM network. The \textit{AirProbe} project on the other hand aims to create a functional sniffer for GSM traffic. So for eavesdropping activities the AirProbe project seems the most logical choice. However, the AirProbe project is still lacking some essential functionality. For one, deciding which type of burst is received is decided by the standard, most likely, division of the broadcast channel, instead of making this decision based on the burst. This means that the results are worse when cell towers use non-standard division of the broadcast channels. Also, currently the AirProbe sniffers can only interpret some types of bursts.
At the moment a lot of development to AirProbe is necessary in order to be able to receive and interpret all of the GSM bursts.

This is mostly a practical issue though, which will be resolved in time.

\section{Countermeasures}\label{sec:cm}
As was already discussed in the introduction, theoretical attacks against GSM are almost as old as the GSM system itself. So the GSM industry has had ample time to prepare for the practical implementations of these attacks.
There are several countermeasures against these attacks, we will discuss the effectiveness of three of these countermeasures here. These are:
\begin{enumerate}
\item Encrypt content using A5/3
\item Use random padding in GSM packets
\item Use UMTS
\end{enumerate}

\subsection{Encrypt content using A5/3}
In Section \ref{sec:th_dec} we already briefly discussed the A5/3 cipher. This cipher has been published from its inception, and as of yet no feasible attack has been found. So using this cipher to encrypt conversations should significantly hamper eavesdropping. 

Indeed, pure eavesdropping will no longer be possible when using A5/3, however it will not improve GSM's security much. This is due to the fact that irrespective of the choice of encryption algorithm, the session key used will be the same. Basically the session key is created based on the secret key, known only to the SIM card and the home network, and a challenge transmitted by the cell tower. This challenge is transmitted in the clear, so an attacker could just record the challenge and an A5/3 encrypted conversation, then at any later time pretend to be a base station to the user and retransmit the challenge. This forces the user's SIM to compute the same session key, which could be broken in an A5/2 connection set up by the fake cell tower, and used to decrypt the conversation. Also, this will not mitigate against MITM attacks.
Finally the GSMA (GSM Association) has been advising the use of A5/3 by providers since 2004, but it seems only a single small provider world-wide has ever made this transition \cite{nohlA51}.

\subsection{Use random padding}
This defensive strategy specifically makes the Kraken attack discussed in Section \ref{sec:prac_dec} harder. It revolves around the fact that the information in GSM packets is padded to a standard length using a standard pattern of ``2b''. Some packets consist almost entirely of padding bits -- the ``cipher mode complete'', the first message a cell phone transmits enciphered to the cell tower, usually has 144 of its 265 bits filled with padding bits -- which gives an attacker a large source of known plain text. However, the length of the information bits is already described in the packet header, making the standard padding pattern redundant.

These padding bits can thus be randomized, and that is exactly what was specified by the ETSI in 2008 \cite{etsi:random}. This would remove a large source of known plaintext for an attacker. Without known plaintext there are no known keystream samples which can be looked up in the Kraken tables.

It is questionable how fast this change will be implemented, however. All the low level GSM processing is done by closed source GSM stacks, so it is unknown whether this change would affect the already deployed equipment. All the mobile handsets in the field can not be updated, so this change can only be made in new phones. Also this change will not completely remove any known plain text from the system. Some messages can still be guessed, such as system information messages, making the attack described in Section \ref{sec:prac_dec} still feasible for longer conversations.

\subsection{Use UMTS}
This is kind of a cop-out, but a method that is at least currently available to quite some users. The successors of GSM -- the 3G systems, mostly UMTS -- offer much better security. Specifically it has mutual authentication between cell tower and mobile phone -- preventing MITM attacks -- and offers stronger encryption, that has seen academic scrutiny.

In order to fully use this added security, a user should deactivate his phone's GSM reception, and solely use UMTS. Otherwise an attacker could force a phone to use GSM, by jamming the UMTS frequencies.
Of course the usability of this solution will depend on the availability of a UMTS network to the specific user, and might have additional data costs.

\section{Conclusions}
Passively eavesdropping on GSM remains pretty hard to do using publicly available hard and software. Theoretically, there are no real constraints in breaking conversation confidentiality in GSM. However, there are still several practical issues preventing a working implementation of a GSM sniffer using freely available hardware.\\

Of the three steps named in Section~\ref{sec:theo} the first -- capturing the GSM signals from the air -- remains the bottleneck. Especially the channel hopping used in GSM networks -- which is not even a security measure -- prevents the correct capture of GSM packets. 

With the current state-of-the-art, the best way to capture GSM data from the air proved the use of an old Nokia phone (3310), which can be put in a debug mode, logging all the GSM bursts it receives. However, this will never reveal bursts that are not meant for this specific phone, and can therefore never be used for eavesdropping. The combination of the USRP, with Gnu Radio and AirProbe currently does not deliver the possibilities needed for eavesdropping.

The release of the rainbow tables and the Kraken tool has made the breaking of the A5/1 encryption much easier. However, this approach does have a few downsides: besides the hard disk size this method also requires perfect samples -- putting additional strain on the capturing process -- and naturally the tables will never give a 100\% chance of finding the key. Still, the current coverage of 22\% of the key space should be workable given enough samples.\\

The practical problems that at the moment prevent a general attack tool can vary with the specific practical situation. Frequency hopping might not be employed by a specific cell tower, or the cell tower transmits on only a few frequencies that lie close together. In fact a cell tower might not even use encryption. In those cases many attacks become much easier, but we do not know if and how many cell towers have such a configuration. During this research all observed cell towers used both frequency hopping and encryption.\\

The same practical problems will probably prevent any general attack tool to be released using the current generation of hardware. It is more likely that a tool will be released that can eavesdrop on some cell towers, though again without publicly available numbers on the configuration of cell towers it is hard to judge how many cell towers are vulnerable. In the mean time the open-source GSM projects have not yet directly worsened the confidentiality of conversations over GSM. Despite many recent claims to the contrary no actual conversation has been captured and decrypted, and it will take a lot of effort before the current problems preventing these attacks are solved.

It is hard to predict how long it will take the current community behind these open-source projects to solve these practical problems. Though reactions from the community seem eager, the recent rate of development in for instance AirProbe do not show much progress.\\

Of the countermeasures that are often referred to by the GSM industry when downplaying the news stories, the most effective one is essentially to by-pass GSM all together and use solely UMTS in stead.

\bibliography{ref}
\bibliographystyle{unsrt}
\end{document}